\documentclass[english,superscriptaddress,twocolumn,aps,prl]{revtex4-2}
\usepackage{graphicx} % Required for inserting images
\usepackage{amsmath}
\usepackage{comment}
\usepackage{xcolor,ulem}
\usepackage{soul}

\begin{document}

\title{Hyperon spin correlation in high-energy heavy-ion collisions}
\author{Xin-Li Sheng}
\email{sheng@fi.infn.it}
\affiliation{Universit\`a di Firenze and INFN Sezione di Firenze,
Via G. Sansone 1, I-50019 Sesto Fiorentino (Florence), Italy}

\author{Xiang-Yu Wu}
\email{xiangyu.wu2@mail.mcgill.ca}
\affiliation{Department of Physics, McGill University, 3600 University Street, Montreal, QC H3A 2T8, Canada}

\author{Dirk H.~Rischke}
\email{drischke@itp.uni-frankfurt.de}
\affiliation{Institut f\"{u}r Theoretische Physik, Johann Wolfgang Goethe-Universit\"{a}t,
Max-von-Laue-Str.~1, D-60438 Frankfurt am Main, Germany}
\affiliation{Helmholtz Research Academy Hesse for FAIR, Campus Riedberg,
Max-von-Laue-Str.~12, D-60438 Frankfurt am Main, Germany}

\author{Xin-Nian Wang}
\email{xnwang@ccnu.edu.cn}
\affiliation{Key Laboratory of Quark and Lepton Physics (MOE) \& Institute of Particle Physics, Central China Normal University, Wuhan 430079, China}
\affiliation{Institut f\"{u}r Theoretische Physik, Johann Wolfgang Goethe-Universit\"{a}t,
Max-von-Laue-Str.~1, D-60438 Frankfurt am Main, Germany}

\begin{abstract}
Recent experimental data show an unexpectedly large spin alignment of $\phi$ mesons in high-energy heavy-ion collisions, which can be explained by short-distance fluctuations of strong-force fields (vector $\phi$ fields) within the constituent-quark model. 
We calculate the hyperon spin correlations within the same model, taking into account hydrodynamic effects and a  $\phi$ field fluctuating in space-time according to a Gaussian distribution. 
The $\Lambda\bar\Lambda$ spin correlation induced by the $\phi$ field is shown to be negative as opposed to that of $\Lambda\Lambda$ or $\bar{\Lambda}\bar{\Lambda}$. 
We thus propose a new net spin-correlation observable as a sensitive probe to separate strong-force effects from hydrodynamic ones. 
With the strength of the field fluctuations extracted from the observed $\phi$ spin alignment, we predict the collision-energy dependence of the hyperon spin correlations and also investigate the dependence of the net spin correlation on azimuthal-angle and rapidity difference. 

\end{abstract}
%\date{July 2024}

\maketitle

\paragraph{\bf Introduction.}

Hot and dense matter produced in non-central heavy-ion collisions, known as the quark-gluon plasma (QGP), carries a large amount of orbital angular momentum (OAM) in the form of a fluid velocity field with nonvanishing vorticity. 
Quarks in the QGP are shown to be globally polarized through the spin-orbital coupling. 
It was predicted that this leads to a global polarization of the final-state hyperons \cite{Liang:2004ph,Voloshin:2004ha,Becattini:2007sr,Becattini:2013fla}. 
Such a phenomenon has been confirmed by the STAR experiment at RHIC \cite{STAR:2017ckg,STAR:2018gyt,STAR:2020xbm,STAR:2021beb} as well as by the HADES experiment at lower energies \cite{HADES:2022enx}, while the global polarization  at LHC energies is close to zero \cite{ALICE:2019onw} as expected. 
Recently, an interesting azimuthal and rapidity dependence of the hyperon polarization has also been observed \cite{STAR:2019erd,ALICE:2021pzu}, which is related to the local structure of the vorticity or the fluid shear in the QGP \cite{Becattini:2013vja,Csernai:2013bqa,Becattini:2015ska,Jiang:2016woz,Pang:2016igs,Becattini:2021suc,Becattini:2021iol,Liu:2021uhn,Fu:2021pok}.

The OAM in heavy-ion collisions was also predicted to lead to the spin alignment of vector mesons \cite{Liang:2004xn}, which was indeed observed recently for $\phi$ mesons \cite{STAR:2022fan}.
However, theoretical predictions for the spin alignment induced by conventional mechanisms such as the fluid vorticity or the magnetic field \cite{Yang:2017sdk,Sheng:2019kmk,Sheng:2022ssp,Wei:2023pdf} are almost two orders of magnitude smaller than what was observed in experiment. 
To account for the disparity, several new mechanisms have been proposed \cite{Sheng:2019kmk,Xia:2020tyd,Muller:2021hpe,Sheng:2022wsy,Sheng:2022ffb,Li:2022vmb,Wagner:2022gza,Kumar:2023ghs,Sheng:2023urn,Dong:2023cng,Sheng:2024kgg,Yang:2024qpy,Chen:2023hnb,Chen:2024afy}. 
Among them is the local strong-force field (SFF) \cite{Sheng:2019kmk,Sheng:2022wsy,Sheng:2022ffb,Sheng:2023urn}, which can polarize quarks through its coupling to the quark spin. 
This field is dominated by fluctuations and thus its mean value over a large space and time scale should be zero. 
As a consequence, it does not contribute to the global polarization of quarks and final-state hyperons. 
However, since the vector meson's spin alignment is related to the spin correlation between the constituent quark and anti-quark \cite{Liang:2004xn,Yang:2017sdk}, the fluctuating SFF can cause the spin alignment if its field correlation is nonzero over the size of the vector meson. 
This model is shown to describe the transverse-momentum dependence of the observed $\phi$ meson spin alignment and predicts the azimuthal-angle or rapidity dependence which can be tested in future experiments \cite{Sheng:2022wsy,Sheng:2023urn}.

Within the constituent-quark model \cite{Fayyazuddin:1994wh,Yang:2017sdk}, the spin of $\Lambda$ ($\bar\Lambda$) is carried by the strange (anti-strange) quark. 
In analogy to the vector-meson spin alignment, one can also expect a spin correlation of hyperons and anti-hyperons that are produced in close proximity on the freeze-out hypersurface due to the fluctuating SFF, if the correlation length of the SFF extends over the size of a hadron. 
In this Letter, we compute spin correlations among $\Lambda$'s and $\bar\Lambda$'s at the freeze-out hypersurface in a (3+1)-D viscous hydrodynamical model, taking into account both the SFF and various hydrodynamic effects such as thermal vorticity, thermal shear, and the spin Hall effect. 
We demonstrate that the difference between ``same-sign" pairs ($\Lambda\Lambda$ or $\bar\Lambda\bar\Lambda$) and ``opposite-sign" pairs ($\Lambda\bar\Lambda$) is a sensitive observable to distinguish the SFF-induced polarization from the conventional hydrodynamic contributions.  
Our predicted spin correlations at RHIC-BES energies are significant and their measurement will shed light on the mechanisms of spin polarization and the nature of the SFF at high temperature.

\paragraph{\bf Spin correlation.}
For a relativistic spin-1/2 particle system, the spin state can be described by a polarization vector $P^{\mu}(x,p)$. 
For a two-particle system composed of particle $1$ and $2$, a straightforward definition of the spin correlation tensor is
\begin{equation}\label{spin-correlation-tensor}
C_{12}^{\mu\nu}(p_{1},p_{2})\equiv\left\langle P_{1}^{\mu}(x_1,p_1)P_{2}^{\nu}(x_2,p_2)\right\rangle\,,
\end{equation}
where $P_{i}^{\mu}$ denotes the polarization vector for particle $i$ with momentum $p_{i}^{\mu}$ at space-time point $x_{i}^{\mu}$. 
Here, $\left\langle \cdots\right\rangle $ includes the event-by-event average as well as the average over space-time points $x_{1}^{\mu}$ and $x_{2}^{\mu}$ weighted by particle densities. 
Since $P_{i}^{\mu}$ is orthogonal to $p_{i}^\mu$, it can be expressed in terms of the 3-components of the spin polarization in the particle's rest frame $(\mathcal{P}_{i,x}^{\text{rest}},\mathcal{P}_{i,y}^{\text{rest}},\mathcal{P}_{i,z}^{\text{rest}})$. 
One then finds that $C_{12}^{\mu\nu}$ contains only 9 independent degrees of freedom, related to the following $3\times3$ matrix
\begin{equation} \label{spin-correlation}
c_{12}^{ab}(p_{1},p_{2})\equiv n_{1,a}^{\mu}(p_{1})n_{2,b}^{\nu}(p_{2})C_{\mu\nu}^{12}(p_{1},p_{2}),\ \ a,b=x,y,z,
\end{equation}
where $n_{i,a}^{\mu}(p_{i})\equiv\Lambda_{\ \nu}^{\mu}(p_{i})\left(0,\,{\bf e}_{a}\right)^{\nu}$ 
with ${\bf e}_{a},\ a=x,y,z$ denoting the unit 3-vector in the direction of the $a$-axis and $\Lambda_{\ \nu}^{\mu}(p_{i})$ being the transformation matrix for the boost from the rest frame of the $i$-th particle to the laboratory frame.
It is straightforward to show that $c_{12}^{ab}=\left\langle \mathcal{P}_{1,a}^{\text{rest}}\mathcal{P}_{2,b}^{\text{rest}}\right\rangle$ is the correlation between the polarizations in the respective rest frames of particles 1 and 2. 
In experiments, $c_{12}^{ab}$ can be extracted from the polar-angle distributions of the decay products if particles 1 and 2 undergo weak decays \cite{Shen:2024buh}, e.g., $\Lambda$ hyperon decays. 
In this work, we concentrate on $\Lambda\Lambda$, $\Lambda\bar{\Lambda}$, and $\bar{\Lambda}\bar{\Lambda}$ pairs, since the multiplicity of $\Lambda$ ($\bar{\Lambda}$) is much larger than other hyperon species.  

Recently the STAR collaboration has observed a significant $\Lambda\bar{\Lambda}$ spin correlation in proton-proton collisions \citep{STAR:2025njp}. 
This correlation is dominated by the hadronization of entangled $s\bar{s}$ pairs produced via virtual gluon splitting, i.e., $g\rightarrow s+\bar{s}\rightarrow \Lambda+\bar{\Lambda}+X$. 
Due to helicity conservation, the spin orientations of $s$ and $\bar{s}$ are enforced to align in the same direction. 
This maximizes the spin correlation of $s\bar{s}$, which is in turn transferred  to the final-state $\Lambda\bar{\Lambda}$ spin correlation \citep{STAR:2025njp,Ellis:1995fc,Ellis:2011kq,Wei:2023pdf,Lin:2025eci}. 
In contrast, in nucleus-nucleus collisions, the $\Lambda\bar\Lambda$ pairs are from the hadronization of the QGP, while their constituent quarks originate from different processes and thus are unlikely to be entangled. 
The spin correlation arises only if both hyperons are polarized by the same external field in the surrounding hot medium. 
Such a kind of correlation corresponds to the “induced spin correlation” in Refs.~\cite{Lv:2024uev,Zhang:2024hyq}. 

\paragraph{\bf Quark and hyperon spin polarization.}

According to the non-relativistic coalescence model \cite{Fayyazuddin:1994wh,Yang:2017sdk,Sheng:2020ghv}, the polarization of a $\Lambda$ baryon is fully  carried by its constituent $s$ quark. We generalize such a relation to the relativistic case by setting
\begin{comment}, ${\bf P}_{\Lambda}({\bf x},{\bf p})=\left\langle {\bf P}_{s}({\bf x}_{s},{\bf p}_{s})\right\rangle _{\Lambda}$, where ${\bf P}_{\Lambda}$ and ${\bf P}_{s}$ are the non-relativistic polarizations of a $\Lambda$ hyperon and an $s$ quark, respectively.
Here, $\left\langle \cdots\right\rangle _{\Lambda}$ denotes an average
over the position ${\bf x}_{s}$ and momentum ${\bf p}_{s}$ of the $s$ quark, weighted by the wave function of the $\Lambda$. 
For a relativistic $\Lambda$ hyperon, we generalize the above relation to the following form
\end{comment}
\begin{equation}
P_{\Lambda}^{\mu}(x,p)\approx P_{s}^{\mu}(x,R_{s}p)\,,\label{eq:relation-between-polarizations}
\end{equation}
where $P_\Lambda^\mu$ and $P_s^\mu$ are polarizations for a $\Lambda$ hyperon and an $s$ quark, respectively, and $R_{s}\equiv m_{s}/m_{\Lambda}$ is the ratio between their invariant masses. 
Here we have neglected the size of the $\Lambda$ and the motion of the $s$ quark relative to the $\Lambda$ hyperon, indicating that the $\Lambda$ hyperon and the constituent $s$ quark have the same position and velocity. 
The polarization of $\bar{\Lambda}$ is related to that of $\bar{s}$ in the same way.
The relation Eq.~(\ref{eq:relation-between-polarizations}) has been previously used in Ref.~\cite{Fu:2021pok} as the ``strange-memory scenario'' to study the local polarization of $\Lambda$.

For a system in local thermodynamical equilibrium, the spin polarization of the $s$ ($\bar{s}$) quark is in general given by 
\begin{equation}\label{quark-polarization}
P^\mu_{s/\bar{s}}(x,p)=P^\mu_\omega+P^\mu_\text{shear}+P^\mu_\text{SHE}+ P^\mu_{\phi}\,,
\end{equation}
where the first three terms on the right-hand-side denote polarizations induced by the thermal vorticity 
$\varpi^{\mu\nu}\equiv(\partial^\nu\beta^\mu-\partial^\mu\beta^\nu)/2$, the thermal shear tensor {$\xi^{\mu\nu}\equiv(\partial^\nu\beta^\mu+\partial^\mu\beta^\nu)/2$}, and the spin Hall effect \cite{Liu:2020dxg}, respectively. 
The last term in Eq.~(\ref{quark-polarization}) is the contribution of the SFF,
\begin{equation}\label{quark-polarization-all}
P^\mu_\phi=\pm\frac{g_\phi}{4mE_pT}\epsilon^{\mu\nu\rho\sigma}F^\phi_{\rho\sigma} p_\nu(1-n_\text{FD}^\pm)\,,
\end{equation}
where $F^\phi_{\mu\nu}$ is the $\phi$ field tensor, $g_\phi$ is the coupling between $\phi$ field and $s$ quark, 
$E_p$ is the quark energy in the center-of-mass frame of the collision, and $n_\text{FD}^\pm \equiv 1/[1+\exp(\beta\cdot p\mp\mu_s/T)]$ is the Fermi-Dirac distribution with $\mu_s$ being the $s$ quark chemical potential and $\beta^\mu\equiv u^\mu/T$ being the thermal velocity. 
Classical electromagnetic fields can also polarize quarks, which are neglected considering that they are sufficiently small at the hadronization stage \cite{McLerran:2013hla,Tuchin:2013apa}. {Feed-down from resonance decays and hadronic afterburner evolution \cite{Becattini:2016gvu,Karpenko:2016jyx} are also neglected.}   

The quantities $u^\mu$, $\mu_s$, $T$, and $F^\phi_{\mu\nu}$ depend on the space-time position $x^\mu$ and thus the total polarization in Eq.~(\ref{quark-polarization}) is locally defined. 
Assuming that the QGP in heavy-ion collisions undergoes a hydrodynamic evolution, the space-time-averaged spin polarization is calculated via the Cooper-Frye formula \cite{Cooper:1974mv,Becattini:2013fla,Becattini:2024uha}. 
For $\Lambda$ ($\bar{\Lambda}$) hyperons with four-momentum $p^\mu$, the latter reads
\begin{equation}\label{Average-polarization}
\left\langle P^\mu_{\Lambda/\bar{\Lambda}}\right\rangle (p)= \sum_{\text{events}}\frac{\int d\Sigma(x)\cdot p\, P^\mu_{\Lambda/\bar{\Lambda}}(x,p)f_{\Lambda/\bar{\Lambda}}(x,p)}{N_\text{event}\int d\Sigma (x)\cdot p\, f_{\Lambda/\bar{\Lambda}}(x,p)}\,,
\end{equation}
where $d\Sigma_\mu(x)$ is the differential volume vector at point $x$ on the freezeout hypersurface and $P^\mu_{\Lambda/\bar{\Lambda}}$ is related to the quark polarization [Eq.~(\ref{quark-polarization})] through Eq.~(\ref{eq:relation-between-polarizations}).
If we also implement the local-equilibrium condition for $\Lambda$ ($\bar{\Lambda}$) hyperons, the distribution $f_{\Lambda/\bar{\Lambda}}$ should be the Fermi-Dirac distribution
\begin{equation}
f_{\Lambda/\bar{\Lambda}}(x,p)=\{1+\exp[\beta(x)\cdot p\mp \mu_B(x)/T(x)]\}^{-1}\,,
\end{equation}
where $\mu_B$ is the baryon chemical potential. 
Analogously, the spin correlation in Eq.~(\ref{spin-correlation-tensor}) is calculated via
\begin{eqnarray}\label{Average-P1-P2}
C^{\mu\nu}_{12}(p_1,p_2)&=&\frac{1}{N_\text{event}}\sum_\text{event}\nonumber\\
&&\hspace{-0.8cm}\times\frac{\int d\Sigma(x_1)\cdot p_1\int d\Sigma(x_2)\cdot p_2\, P^\mu_1\, P^\nu_2\, f_1\,f_2\,}{\left[\int d\Sigma(x_1)\cdot p_1\, f_1\right]\left[\int d\Sigma(x_2)\cdot p_2\, f_2\right]}\,,
\end{eqnarray}
where $P_i^\mu,\,f_i,\,i=1,2$ are functions of $x_i^\mu$ and $p_i^\mu$, and particles $1$, $2$ denote $\Lambda$ or $\bar{\Lambda}$ hyperons.

\paragraph{\bf Numerical setup.}

In this work, we employ the CLVisc hydrodynamic model \cite{Pang:2018zzo,Wu:2021fjf,Yi:2021ryh,Wu:2022mkr} to generate the hadronization hypersurface and the corresponding hydrodynamic fields. To calculate the spin correlation induced by hydrodynamic effects, i.e., $P^\mu_\omega$, $P^\mu_\text{shear}$, and $P^\mu_\text{SHE}$, we first extract space-time-dependent values of $\varpi_{\mu\nu}$, $\xi_{\mu\nu}$, $T$, and $\mu_s$ for 5000 event-by-event simulations with fluctuating SMASH initial conditions \cite{Wu:2022mkr,Fu:2021pok,Fu:2020oxj,Becattini:2021iol}. The hadronization hypersurface is determined by the local energy density reaching $0.4 \text{ GeV}/\text{fm}^3$. Then, the spin correlation in absence of SFFs is evaluated using Eqs.~(\ref{spin-correlation-tensor}), (\ref{spin-correlation}), (\ref{eq:relation-between-polarizations}), and (\ref{Average-P1-P2}).

Furthermore, we assume that fluctuations of the vector $\phi$ field have very weak correlations to hydrodynamic quantities. This implies that we can neglect the correlation between the polarization induced by hydrodynamic effects and that induced by the $\phi$ field. 
We assume that the $\phi$ field tensor $F^\phi_{\mu\nu}$ fluctuates in space-time while the field-field correlation is parameterized as 
\begin{align}
&g_{\phi}^{2}F_{ij}^{\phi}(x_1)F_{ij}^{\phi}(x_2)/[T(x_1)T(x_2)] \nonumber\\
&=\begin{cases}
F_{T}^{2}G(x_1-x_2), & (i,j)=(0,1),\,(0,2),\,(2,3),\,(3,1)\,,\\
F_{z}^{2}G(x_1-x_2), & (i,j)=(0,3),\,(1,2)\,,
\end{cases}\,\label{eq:field-correlations}
\end{align}
while other unlisted components are zero.
The long-distance correlation is suppressed by a Gaussian function
\begin{equation}
G(x_1-x_2)\equiv\text{exp}\left[-\frac{(t_1-t_2)^{2}}{\sigma_{t}^{2}}-\frac{({\bf x}_1-{\bf x}_2)^{2}}{\sigma_{x}^{2}}\right]\,,\label{eq:Gaussian-function}
\end{equation}
where $\sigma_{t}$ and $\sigma_{x}$ are typical temporal and spatial scales of the field correlation. 
In our calculation we take $\sigma_{t,x}=0.6\,(\pm0.3)$ fm, which is comparable to the radius of light mesons, {with a more detailed justification provided in the supplemental material \cite{ParticleDataGroup:2024cfk,Hawes:1998bz,Blaizot:2001nr,Gelis:2010nm,campostrini1986g,DIGIACOMO2002319}}. 
The parameters $F_T^2$ and $F_z^2$ are local fluctuations of transverse and longitudinal fields, which are extracted in Ref.~\cite{Sheng:2022wsy} by fitting the $\phi$ meson spin-alignment data from the STAR experiment. 
They can be effectively parametrized by the following analytical form \cite{Sheng:2022wsy} as a function of the colliding energy $\sqrt{s_{\text{NN}}}$,
\begin{equation}
\ln(F_{T,z}^{2}/m_{\pi}^{2})=a_{T,z}-b_{T,z}\ln(\sqrt{s_{\text{NN}}}/\text{GeV})\,,
\end{equation}
with $m_{\pi}=0.138$ GeV, $a_{T}=3.90\pm1.11$, $b_{T}=0.924\pm0.234$, $a_{z}=3.33\pm0.917$, and $b_{z}=0.760\pm0.189$. 
Here the intrinsic anisotropy of the QGP induced by a faster longitudinal expansion relative to the transverse expansion is taken into account, while the difference between the in-plane direction and the out-of-plane direction is neglected. 
Since the $\phi$ meson spin alignment is more strongly influenced by the relative motion of the meson to the background than by the transverse anisotropy of the QGP, as pointed out in Ref.~\cite{Sheng:2023urn}, this setup should capture the dominant behavior of the $\Lambda$ spin correlation. 
{The spin correlation induced by SFFs is evaluated within a hydrodynamic simulation with the event-averaged initial condition because here the double integral over the freeze-out hypersurface in Eq.~\eqref{Average-P1-P2} does not factorize, and thus an event-by-event calculation would exceed present computation-time resources.}

\begin{figure}
\includegraphics[width=0.9\linewidth]{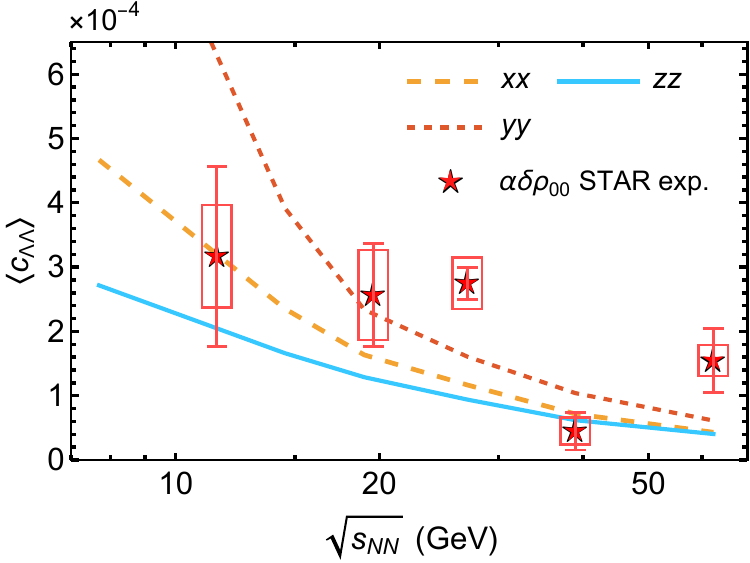}
\caption{\label{fig:Spin-correlation-energy} The collision-energy dependence of the $\Lambda\Lambda$ spin correlation, compared to  experimental data \cite{STAR:2022fan} (red stars with error bars) for the $\phi$ meson global spin alignment $\delta\rho_{00}\equiv \rho_{00}-1/3$. The $xx$, $yy$, and $zz$ components of $c^{ab}_{\Lambda\Lambda}$ are denoted by the orange long-dashed line, the red short-dashed line, and the cyan solid line, respectively. The global spin alignment is rescaled by a factor $\alpha=0.01$ such that it has the same order of magnitude as the spin correlation.}
\end{figure}

\paragraph{\bf Numerical results.}
We present in Fig.~\ref{fig:Spin-correlation-energy} the diagonal components of the $\Lambda\Lambda$ spin correlation (\ref{spin-correlation}) as functions of the colliding energy. 
The calculations are carried out for Au+Au collisions at $\sqrt{s_\text{NN}}=7.7$, $14.5$, $27$, $39$, and $62.4$ GeV, within the centrality class of $20-50\%$, and are averaged over hyperons at central rapidity $|Y|<1$ with transverse momentum $p_{T}\in(0.5,\,3)$ GeV/$c$. 
The spin correlation is stronger at lower energies, showing the same behavior as the $\phi$ meson spin alignment \cite{STAR:2022fan}. 
For the $xx$ and $zz$ components, the correlation is governed by SFF fluctuations, which are parameterized as decreasing functions of the collision energy in order to fit the experimental data for $\rho_{00}$ of the $\phi$ meson. 
Furthermore, the SFF has nearly the same contribution to $c_{\Lambda\Lambda}^{xx}$ and $c_{\Lambda\Lambda}^{yy}$. 
The splitting between them in Fig.~\ref{fig:Spin-correlation-energy} is mainly a consequence of the global spin polarization of quark and anti-quark induced by the vorticity field. 
The $zz$ component of the spin correlation appears to be smaller than the transverse components, reflecting the asymmetry between the fireball's longitudinal and transverse expansion. 

\begin{figure}
\includegraphics[width=0.9\linewidth]{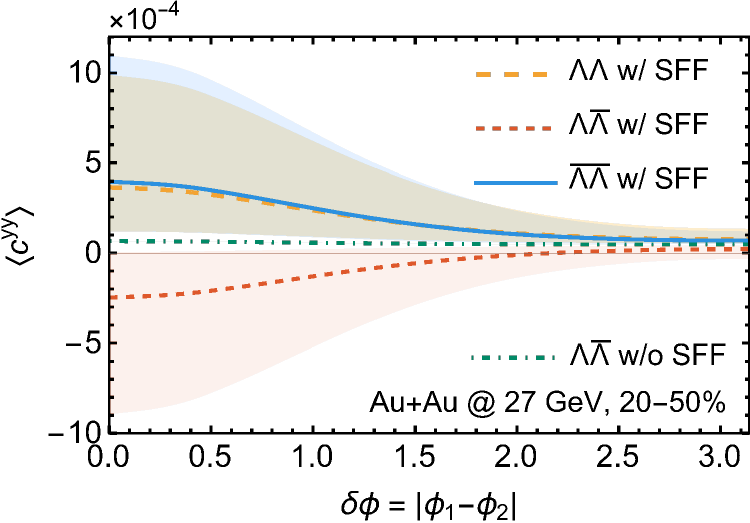}

\caption{\label{fig:Spin-correlation-phi}
The spin correlation component $c_{12}^{yy}$ as a function of the azimuthal-angle difference between particles $1$ and $2$ {in the range $|Y|<1$, $p_T\in(0.5,\,3)$ GeV/c}. Correlations in the presence of SFF for $\Lambda\Lambda$, $\Lambda\bar{\Lambda}$, and $\bar\Lambda\bar\Lambda$ are denoted by orange long-dashed, red short-dashed, and cyan solid lines, respectively. The contribution from hydrodynamic fields to $c_{\Lambda\bar\Lambda}^{yy}$ is shown by the green dash-dotted line, while $c_{\Lambda\Lambda}^{yy}$ and $c_{\bar\Lambda\bar\Lambda}^{yy}$ in absence of SFF are nearly identical with $c_{\Lambda\bar\Lambda}^{yy}$ and thus are not shown in this figure. {The lines are calculated with $\sigma_{t,x}=0.6$ fm, while the uncertainty bounds shown by the shaded areas are determined by $\sigma_{t,x}=0.3,\,0.9$ fm, respectively.}}
\end{figure}

Figure \ref{fig:Spin-correlation-phi} displays the $yy$ component of the spin correlation and its uncertainty.
The absolute values of the spin correlations reach their maximum at $\delta\phi=0$ and decrease towards zero as $\delta\phi\rightarrow\pi$, suggesting that two hyperons are more (less) correlated if they move in the same (opposite) direction. 
A slight difference is observed between the spin correlation of $\Lambda\Lambda$ and that of $\bar{\Lambda}\bar{\Lambda}$ as a consequence of different space-time distributions of $\Lambda$ and $\bar{\Lambda}$ when the baryon chemical potential is nonzero. 
Even in the absence of SFF fluctuations (shown as the green dash-dotted line), there still exist nonzero spin correlations, which are mainly attributed to the global rotation of the QGP. 
When $\delta\phi=0$,
spin correlations between hyperons are also dominated by SFF, much like the vector-meson spin alignment. 
When $\delta\phi=\pi$, the hydrodynamic fields, especially the vorticity field, become the dominant source for the spin correlation, while the magnitude is approximately equal to the product of $\Lambda$'s and $\bar{\Lambda}$'s global polarizations, $P_\Lambda P_{\bar{\Lambda}}\sim \mathcal{O}(10^{-4})$. 
{The shaded areas in Fig. \ref{fig:Spin-correlation-phi} show the variations due to the uncertainty of the correlation length $\sigma_{t,x}$. A larger $\sigma_{t,x}$ corresponds to larger absolute values for the spin correlation.} 

\begin{figure}

\includegraphics[width=0.9\linewidth]{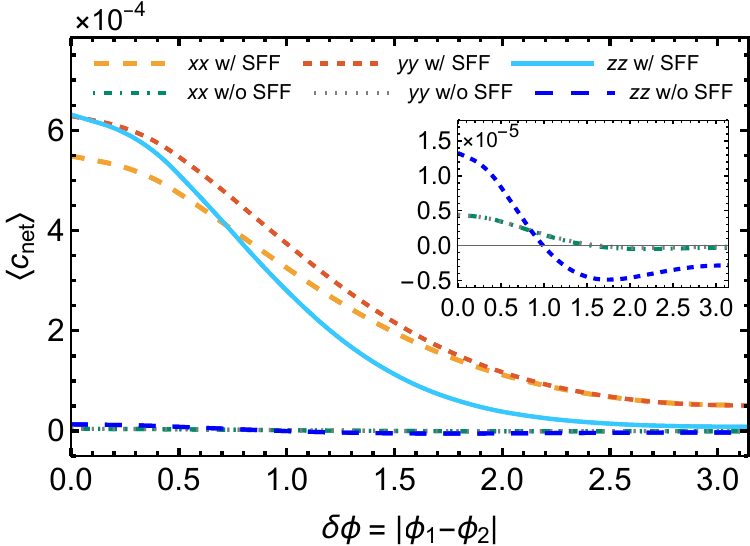}

\caption{\label{fig:net-Spin-correlation-phi} Components of the net spin correlation $c^{ab}_\text{net}$ as functions of the azimuthal-angle difference in the range $|Y|<1$, $p_T\in(0.5,\,3)$ GeV/c for Au+Au collisions at $\sqrt{s_\text{NN}}=27$ GeV. Results with the SFF are shown by the orange, red, and cyan lines, while the other lines are results in absence of the SFF.}
\end{figure}

\begin{figure}

\includegraphics[width=0.9\linewidth]{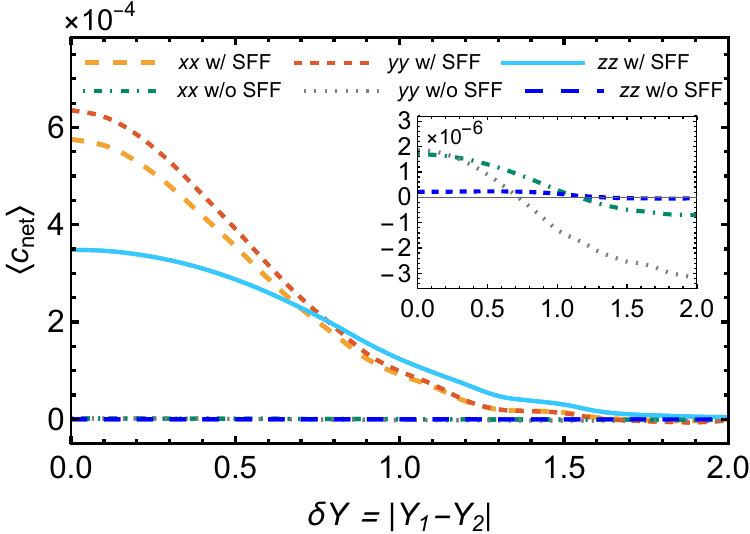}

\caption{\label{fig:net-Spin-correlation-y} Components of the net spin correlation $c^{ab}_\text{net}$ as functions of the rapidity difference. Computational conditions and notations are the same as in Fig.~\ref{fig:net-Spin-correlation-phi}.}
\end{figure}

A key unique feature of the hyperon spin correlation caused by the SFF is that the spin correlation of $\Lambda\Lambda$ and $\bar\Lambda\bar\Lambda$ has the opposite sign as the $\Lambda\bar\Lambda$ spin correlation, as shown in Fig.~\ref{fig:Spin-correlation-phi}.
In order to separate the SFF contribution from others, we introduce the following net spin correlation,
\begin{equation}
c_{\text{net}}^{ab}\equiv \frac{1}{2}(c^{ab}_{\Lambda\Lambda}+c^{ab}_{\bar\Lambda\bar\Lambda})-c^{ab}_{\Lambda\bar\Lambda}\,,
\end{equation}
which denotes the difference between the ``same-sign” pairs ($\Lambda\Lambda$ or $\bar\Lambda\bar\Lambda$) and the ``opposite-sign" pairs ($\Lambda\bar\Lambda$). 
We show in Figs.~\ref{fig:net-Spin-correlation-phi} and \ref{fig:net-Spin-correlation-y} its dependence on the azimuthal-angle difference and the rapidity difference between the two hyperons. 
When the SFF is taken into account, the $xx$, $yy$, and $zz$ components of $c_\text{net}$ peak at $\delta\phi=0$ or $\delta Y=0$ and approach  zero when $\delta\phi$ or $\delta Y$ increase.
%, as indicated by the orange, red, and cyan lines. 
In the absence of SFF, $c_\text{net}$ becomes 1-2 orders of magnitude smaller than in the case with SFF, shown in the insets in Figs.~\ref{fig:net-Spin-correlation-phi} and \ref{fig:net-Spin-correlation-y}. 
This behavior arises because the SFF has an opposite effect on the polarizations of $\Lambda$ and $\bar{\Lambda}$, as described by Eq.~(\ref{quark-polarization-all}), leading to opposite contributions to the spin correlations of $\Lambda\Lambda$ and $\Lambda\bar\Lambda$. 
In contrast, the polarizations induced by the thermal vorticity and the thermal shear tensor are identical for $\Lambda$ and $\bar{\Lambda}$, resulting in nearly the same values of $c_{\Lambda\Lambda}^{yy}$, $c_{\bar{\Lambda}\bar{\Lambda}}^{yy}$, and $c_{\Lambda\bar{\Lambda}}^{yy}$ when the SFF is switched off. 
The spin Hall effect could also result in a nonzero $c_\text{net}$, but its magnitude is much smaller than the contribution of SFF. 
Therefore $c_\text{net}$ is shown to be sensitive to the SFF and is almost independent from the other hydrodynamic fields.

The dependence of the net spin correlation on $\delta \phi$ or $\delta Y$ is a consequence of the position-momentum correlation of hyperons. 
Due to the transverse expansion of the QGP, the mean transverse position of hyperons emitted from the freeze-out hypersurface is approximately aligned with the transverse momentum, $\left\langle {\bf x}_T\right\rangle \propto \hat{\bf p}_T$. 
Similarly, the longitudinal expansion leads to a nearly linear dependence of $\left\langle z\right\rangle$ on the rapidity. 
Given such significant position-momentum correlations, a larger separation of two hyperons in momentum space corresponds to a larger distance between their production positions, which leads to weaker field-field and spin-spin correlations according to Eqs.~(\ref{Average-P1-P2}) and (\ref{eq:field-correlations}). 
The $\delta \phi$ or $\delta Y$ dependence is similar to the $\Lambda\bar\Lambda$ spin correlation observed in p+p collisions \cite{STAR:2025njp}, but arises from a different mechanism. 
In p+p collisions (or for high-momentum hyperons in A+A collisions \cite{Li:2023qgj,Huang:2024awn}), it is dominated by the parton fragmentation, while the correlation in this study is dominated by interactions between constituent quarks and the fluctuating SFF.

\paragraph{\bf Summary and Discussion.}

Based on the constituent-quark coalescence model, we have investigated the spin correlation between $\Lambda$ ($\bar\Lambda$) hyperons using the CLVisc hydrodynamic model with SMASH initial conditions. 
We considered the spin polarization due to both the hydrodynamic fluid evolution and the SFF with a short-distance Gaussian correlation. 
Due to opposite contributions of the SFF to the $\Lambda$ and $\bar\Lambda$ polarization, the spin correlation induced by the SFF fluctuation is negative for $\Lambda\bar\Lambda$ while it is positive for $\Lambda\Lambda$ and $\bar\Lambda\bar\Lambda$. 
We thus proposed the net spin correlation, which describes the difference between ``same-sign'' pairs and ``opposite-sign'' pairs, as a unique signature of spin polarization induced by SFF in heavy-ion collisions. 

Given the SFF fluctuation constrained by the experimental data on the $\phi$ meson spin alignment, we have provided quantitative predictions on the relative momentum dependence ($\delta\phi$ and $\delta Y$) of the net spin correlations of hyperons, which all decrease with their separation in momentum space. 
We also predicted the dependence of different spin-correlation components on the collision energy, which shows a pattern similar to the $\phi$ meson spin alignment. 
Provided that the typical space-time scale of the SFF fluctuation is comparable to the size of a light meson, $\sigma_{t,x}=0.6$ fm, the net spin correlation is $\mathcal{O}(10^{-4})$. Larger values of $\sigma_t$ and $\sigma_x$ result in stronger spin correlations.
{A simple estimate for the required statistical precision shows that at least $10^8$ pairs are needed for measuring such a tiny effect, which can be achieved in the BES-II program at RHIC.}

These predictions provide a perfect opportunity to test the mechanism of spin polarization in the QGP and constraints on the scale of the SFF fluctuation in high-energy heavy-ion collisions. 
{It will provide new information about the short-range correlation in the QGP and the non-perturbative behaviour of strongly interacting matter at high temperature and density.}
For simplicity, we have neglected the internal motion of the $s$ quark inside the hyperon in the constituent-quark coalescence model. 
Additionally, the classical electromagnetic fields are not considered, as they should be weak at the freeze-out stage of heavy-ion collisions. 
These effects, along with further refinements, can be addressed in future studies.  

\paragraph{\bf Acknowledgments.} 

The authors thank Diyu Shen and Aihong Tang for enlightening discussions. 
This work is supported in part by the Italian Ministry of University and Research, project PRIN2022``Advanced probes of the Quark Gluon Plasma", Next Generation EU, Mission 4 Component 1, X.-Y.W.~is supported in part by the Natural Sciences and Engineering Research Council of Canada (NSERC) [SAPIN-2020-00048 and SAPIN-2024-00026] and in part by US National Science Foundation (NSF) under grant number OAC2004571. D.H.R.~is supported by the \textit{Deutsche Forschungsgemeinschaft} (DFG, German Research Foundation) through CRC-TR 211 \textit{``Strong-interaction matter under extreme conditions''} - project number 315477589 - TRR 211.
X.-N.W. is supported by NSFC under Grant No.~1193507, the Guangdong MPBAR with No.~2020B0301030008 and by the Alexander von Humboldt Foundation through a Humboldt Research Award. The data that support the findings of this article are openly available \cite{sheng_2025_17938737}.

\bibliographystyle{apsrev}
\bibliography{Spin-correlation}

% -------------------------------------------------------
% Supplemental Material starts here
% -------------------------------------------------------

\clearpage\newpage

$\ $

$\ $

\onecolumngrid  % optional: switch to one-column for SM

\begin{center}
\textbf{\large Supplemental Material for\\
``Hyperon spin correlation in high-energy heavy-ion collisions''}
\end{center}

\setcounter{section}{0}
\setcounter{figure}{0}
\setcounter{table}{0}
\setcounter{enumiv}{0}
\setcounter{equation}{0}

\renewcommand{\thesection}{S\arabic{section}}
\renewcommand{\thefigure}{S\arabic{figure}}
\renewcommand{\thetable}{S\arabic{table}}
\renewcommand{\theenumiv}{S\arabic{enumiv}} 
\renewcommand{\theequation}{S\arabic{equation}}

\section{Measuring spin correlation in experiments}
In experiments, the spin polarization of the $\Lambda$ hyperon is extracted from its weak decay channel $\Lambda\rightarrow p+\pi^-$. The angular distribution of the daughter proton follows
\begin{equation}
\frac{dN}{d\Omega^\ast}=\frac{1}{4\pi}\left(1+\alpha_\Lambda \frac{\boldsymbol{\mathcal{P}}^\text{rest}_\Lambda\cdot {\bf p}^\ast}{|{\bf p}^\ast|} \right)\,,
\end{equation}
where $\boldsymbol{\mathcal{P}}^\text{rest}_\Lambda$ is the spin-polarization vector of $\Lambda$, ${\bf p}^\ast$ is the polar angle of the proton momentum in the rest frame of the $\Lambda$, and $\alpha_\Lambda$ denotes the decay constant. This relation allows one to measure the polarization through 
\begin{equation}
\boldsymbol{\mathcal{P}}^\text{rest}_\Lambda=\frac{3}{\alpha_\Lambda}\left\langle \hat{\bf p}^\ast\right\rangle\,.
\end{equation}
where $\hat{\bf p}^\ast\equiv{\bf p}^\ast/|{\bf p}^\ast|$ is the unit vector along the direction of ${\bf p}^\ast$.  In analogy to this approach, the hyperon spin correlation introduced in Eq.~(\ref{spin-correlation}) is measured from the decay products via
\begin{equation}
c^{ab}_{12}=\frac{9}{\alpha_1\alpha_2}\left\langle \cos\theta_{1,a} \cos\theta_{2,b}\right\rangle\,,
\end{equation}
where $1,2$ refer to $\Lambda$ or $\bar{\Lambda}$, and $a,b=x,y,z$ denote three spatial directions. Here, $\cos\theta_{1,a}\equiv\hat{\bf p}_1^\ast\cdot {\bf e}_a$ and $\cos\theta_{2,b}\equiv\hat{\bf p}_2^\ast\cdot {\bf e}_b$ with $\hat{\bf p}^\ast_{i}$ 
being the unit vector of the daughter-proton's momentum from the decay of the $i$-th hyperon, and ${\bf e}_{x,y,z}$ the unit vectors along the $x,y,z$ axes. The parameters $\alpha_1$ and $\alpha_2$ are the decay constants for hyperons 1 and 2, respectively.

\section{Spin polarization of quark/antiquark}

In the quark-gluon plasma described by a relativistic fluid, quarks and antiquarks can be polarized by various sources. In this work, we take into account the contributions from the thermal vorticity, the thermal shear tensor, the spin Hall effect, and the SFF. The average polarization as a function of phase-space coordinates is defined as
\begin{equation}
P^\mu(x,p)\equiv n^\mu(x,p) \frac{f_+(x,p)-f_-(x,p)}{f_+(x,p)+f_-(x,p)}\;,
\end{equation}
where $n^\mu(x,p)$ denotes the unit vector along the polarization direction for a particle (or antiparticle) with momentum $p$ at position $x$, and $f_\pm(x,p)$ are the number densities of the particle (or antiparticle) with spin projection $\pm1/2$ along $n^\mu$. As shown in Eq.~(\ref{quark-polarization}),  the average polarization can be decomposed into several contributions, where
\begin{align}\label{eq:explicit-polarization}
P_\omega^\mu(x,p)&=-\frac{1}{4m}\epsilon^{\mu\nu\rho\sigma}p_\nu\varpi_{\rho\sigma}\;, \nonumber\\
P_\text{shear}^\mu(x,p)&=-\frac{1}{2m(u\cdot p)}\epsilon^{\mu\nu\rho\sigma}p_\nu u_\rho\xi_{\sigma\lambda}p^\lambda\;, \nonumber\\
P_\text{SHE}^\mu(x,p)&=\pm\frac{1}{2m (u\cdot p)}\epsilon^{\mu\nu\rho\sigma}p_\nu u_\rho\partial_\sigma\frac{\mu_s}{T}\;,
\end{align}
with $P_\phi^\mu$ given in Eq.~(\ref{quark-polarization-all}). Here, $p^\mu=(\sqrt{{\bf p}^2+m^2},{\bf p})$ is the four-momentum for the considered particle (or antiparticle), while $u\cdot p$ is its energy in the frame comoving with the fluid. The explicit expression for $P_\omega^\mu(x,p)$ can be found in the review \cite{Becattini:2024uha}, where $S^\mu$ in this reference corresponds to $P^\mu/2$ in our notation. The shear-induced polarization has been formulated differently in different papers \cite{Becattini:2021suc,Becattini:2021iol,Liu:2021uhn,Fu:2021pok}. In this work, we take the frame vector in $P_\text{shear}^\mu$ to be $u^\mu$, which is consistent with the formula in Ref. \cite{Fu:2021pok}.

\section{Estimate of the correlation length}

When calculating the spin correlation induced by SFFs, we have employed a Gaussian smearing function (\ref{eq:Gaussian-function}) to suppress long-distance field correlations. This introduces $\sigma_t$ and $\sigma_x$ as the typical temporal and spatial scales, which can also be interpreted as the typical correlation length of the SFF (the $\phi$ field in this work). From the theoretical side, there is no precise model input for $\sigma_{t,x}$, so here we provide several estimates for the correlation length. The SFF interaction provides an effective description of the strong interaction between particles at the freeze-out stage, which lies intermediate between the hadronic interaction at low temperature and the gluonic interaction at high temperature. It is therefore natural to expect that $\sigma_{t,x}$ falls between the correlation length in a hadron gas and that of pure QCD. Several possible estimates are:
\begin{itemize}
    \item[1.] In QCD at low energies, the hadronic interactions are predominantly mediated by mesons. Since mesons have a nonzero spatial extent rather than being point-like particles, it is reasonable to assume that the SFFs also extend over the meson's spatial extent. The corresponding correlation length can thus be estimated by the charge radii of mesons: 0.66 fm for pions \cite{ParticleDataGroup:2024cfk}, 0.56 fm for kaons \cite{ParticleDataGroup:2024cfk}, and 0.61 for $\rho$ mesons \cite{Hawes:1998bz}.
    \item[2.] Alternatively, one can estimate the correlation length using the typical interaction scale, i.e., the Compton wavelength $\lambda\sim1/m$. Taking $m$ as the vacuum masses of pion, kaon, or $\phi$ meson, the related scales are 1.4 fm, 0.4 fm, and 0.2 fm, respectively.  
    \item[3.] In hot QCD matter where the strong interaction is mediated by gluon fields, the correlation length of color electric fields can be estimated via the Debye screening length \cite{Blaizot:2001nr}, $\lambda_D=1/m_D\sim1/(gT)$. Taking the freeze-out temperature $T\approx160$ MeV and the strong coupling constant $g\approx 2$, the Debye screening length is $\lambda_D\approx0.62$ fm.
    \item[4.] In the CGC/Glasma framework, the transverse correlation length of the color fields is parametrized by the inverse saturation scale, $\lambda\sim1/Q_s$ \cite{Gelis:2010nm}. The typical value $Q_s\sim$ 1 GeV corresponds to $\lambda\sim$ 0.20 fm.
    \item[5.] According to lattice-QCD calculations, the correlation length of the gluon field is around 0.2 -- 0.3 fm, or even shorter \cite{campostrini1986g,DIGIACOMO2002319}.
\end{itemize}

In our work, the spin correlation is calculated at the freeze-out stage of the collision, which lies intermediate between a hadron gas and a pure QCD matter. Therefore, the correlation length of SFF used here is expected to be neither as short as that of gluon field nor as long as the pion Compton wavelength. Based on the above considerations, we adopt a conservative estimate for the correlation length of $\sigma_{t,x}=0.6$ fm, with an uncertainty of $0.3$ fm.

\section{$\delta p_T$-dependence of spin correlation}

\begin{figure}
\includegraphics[width=0.45\linewidth]{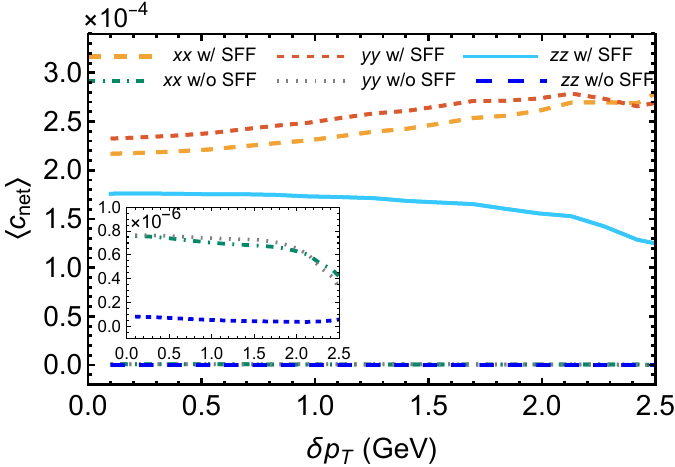}

\caption{\label{fig:Spin-correlation-pT} The net spin correlation as a function of $\delta p_T\equiv|p_{T,1}-p_{T,2}|$. Computational conditions are the same as in Fig.~\ref{fig:net-Spin-correlation-phi}.}
\end{figure}

We show the net spin correlation as a function of the transverse-momentum difference in Fig.~\ref{fig:Spin-correlation-pT}, where $\delta p_T\equiv|p_{T,1}-p_{T,2}|$. In the presence of SFFs, the $xx$ and $yy$ components of the net spin correlation increase with $\delta p_T$, while the $zz$ component decreases. Contributions from hydrodynamic fields are found to be two orders of magnitude smaller than those from the SFFs, as illustrated in the inset of Fig.~\ref{fig:Spin-correlation-pT}.

\section{Effect of spin Hall effect to spin correlation}

\begin{figure}
\includegraphics[width=0.45\linewidth]{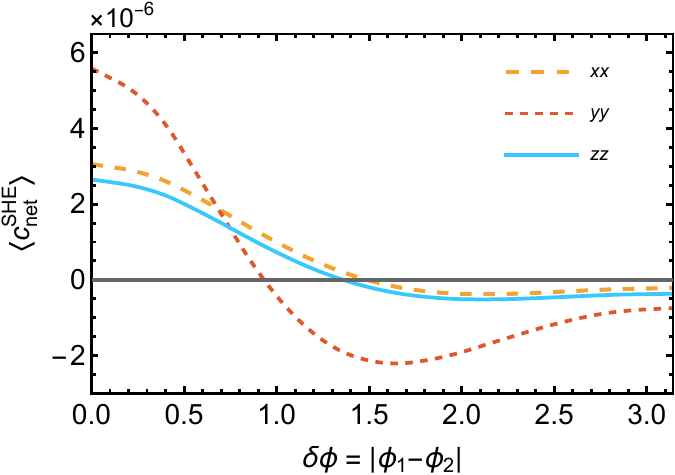}

\caption{\label{fig:Spin-correlation-phi-SHE} The contribution of SHE to the net spin correlation as a function of $\delta\phi$. Computational conditions are the same as in Fig.~\ref{fig:net-Spin-correlation-phi}.}
\end{figure}

\begin{figure}
\includegraphics[width=0.45\linewidth]{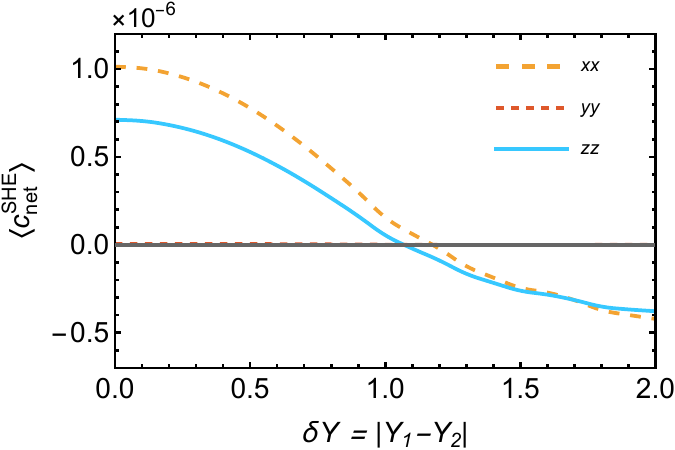}

\caption{\label{fig:Spin-correlation-y-SHE} The contribution of SHE to the net spin correlation as a function of $\delta Y$. Computational conditions are the same as in Fig.~\ref{fig:net-Spin-correlation-phi}.}
\end{figure}

The spin Hall effect, originating form the gradient of $\mu_s/T$, has opposite contributions to the polarization of quark and antiquark, as shown in Eq. (\ref{eq:explicit-polarization}). Consequently, it could induce $c_\text{net}$ in a similar way as the SFFs. We show in Figs.~\ref{fig:Spin-correlation-phi-SHE} and \ref{fig:Spin-correlation-y-SHE} its contributions to $c_\text{net}$ as functions of $\delta\phi$ and $\delta Y$, respectively. The resulting correlations are of order $\mathcal{O}(10^{-6})$, which is significantly smaller than those induced by the SFFs.

\end{document}